\documentclass[conference,10pt]{IEEEtran}

\usepackage{amsmath}
\usepackage{graphicx,subfigure,comment}
\usepackage[table]{xcolor}
\usepackage{url} 
\usepackage{dsfont}
\usepackage{cite}
\usepackage{tabularx}
\usepackage{dblfloatfix}

\newcommand{\Fig}[1]{Fig.~\ref{fig:#1}}

\newcommand{\Sec}[1]{Sec.~\ref{sec:#1}}
\newcommand{\Tab}[1]{Tab.~\ref{tab:#1}}
\newcommand{\Eq}[1]{(\ref{eq:#1})}

\begin{document}

\title{Flexible Parallel Learning in Edge Scenarios:\\Communication, Computational and Energy Cost}

\author{\IEEEauthorblockN{Francesco Malandrino}
\IEEEauthorblockA{CNR-IEIIT and CNIT\\
Torino, Italy}
\and
\IEEEauthorblockN{Carla Fabiana Chiasserini}
\IEEEauthorblockA{Politecnico di Torino, CNR-IEIIT and CNIT\\
Torino, Italy}
} 

\maketitle

\begin{abstract}
Traditionally, distributed machine learning takes the guise of (i) different nodes training the same model (as in federated learning), 
or (ii) one model being split among multiple nodes (as in distributed stochastic gradient descent). 
In this work, we highlight how fog- and IoT-based scenarios often require combining both approaches, 
and we present a framework for {\em flexible parallel learning (FPL)}, achieving both data and model parallelism. 
Further, we investigate how different ways of distributing and parallelizing learning tasks across the participating nodes 
result in different computation, communication, and energy costs. Our experiments, carried out using state-of-the-art 
deep-network architectures and large-scale datasets, confirm that FPL allows for  
an excellent trade-off among computational (hence energy) cost, communication overhead, and 
learning performance.
\end{abstract}


\section{Introduction}
\label{sec:intro}

\begin{figure}
\centering
\vspace{2mm} 
\includegraphics[width=.48\columnwidth]{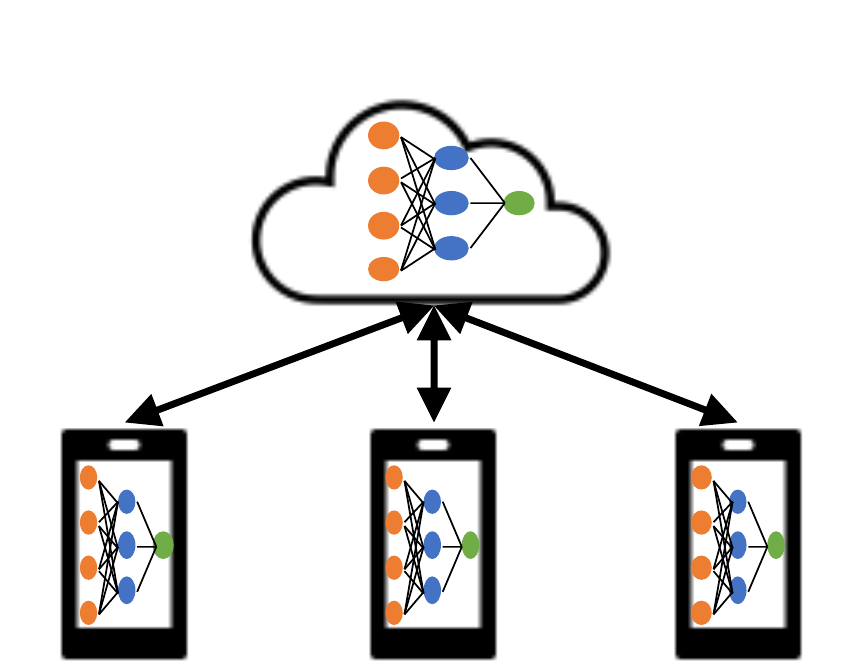}
\includegraphics[width=.48\columnwidth]{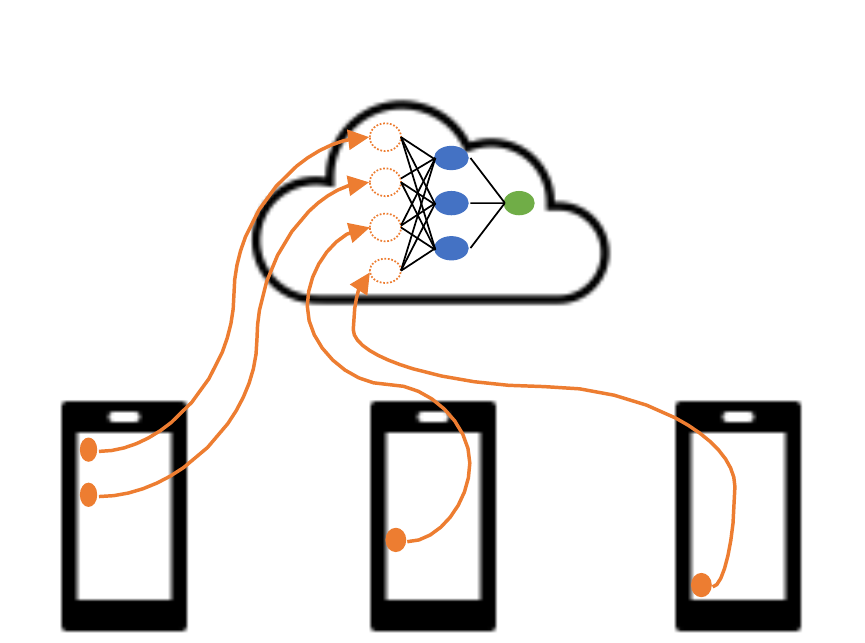}
\caption{
Two of the main existing approaches to distributed supervised learning: FL (left) and D-SGD (right).
\label{fig:approaches}
} 
\end{figure}

The emerging pervasiveness of machine learning (ML) and the fact that data generated by people, machines and sensors is expected to soon 
amount to 850\,ZB~\cite{cisco} have led to an increasing adoption of distributed ML, 
involving multiple network nodes.
Reasons to adopt this new paradigm include the ability 
to leverage more computational and energy resources, and the possibility of exploiting local data without 
disclosing it~\cite{konen2015federatedOptimization,kang2020reliable} or transporting it to far-away data centers. 
This is particularly important for the training of ML models, especially in the case of the popular 
deep neural networks~\cite{dean2012large} (DNNs). DNNs are indeed composed 
of many layers, requiring large amounts of data to set the many parameters they are composed of and 
take a  high toll in terms of computing resource and energy  consumption. 
There are currently two main approaches to distributed supervised learning: 
federated learning (FL)~\cite{konen2015federatedOptimization,kang2020reliable,wang2019adaptive} and 
distributed stochastic gradient descent
(D-SGD)~\cite{gemulla2011large}, both depicted in \Fig{approaches}.
Under both FL and D-SGD, the total training time includes both
(i) local computation, and (ii) network delay.

Under FL, all nodes share the same DNN architecture and  
each node trains its DNN with local data. 
After one or more local training epochs, 
local parameters are sent to a centralized learning server, 
which is in charge of combining them  (by averaging them, or through more complex strategies~\cite{noi-commag}), 
and sending the results back to the learning nodes. The main appeal of FL is its ability to exploit local data for learning, 
without sharing it, which is especially important for private and/or potentially sensitive data. 
Further, FL is suitable for scenarios where learning nodes, such as mobile user devices, can appear or disappear in an unpredictable manner,
including network resource scheduling~\cite{jiang2019deep}
and the management of drone-powered MEC systems~\cite{jiang2020ai}.
As multiple learning nodes train different instances of the same model on different sets of data, 
FL is said to implement {\em data parallelism}~\cite{krizhevsky2014one}.

On the contrary, the D-SGD paradigm allows for {\em splitting} a given DNN architecture among learning nodes, enabling each node to run only 
a part of the DNN. 
Nodes communicate with each other during each learning iteration (both forward- and back-passes), 
exchanging information on the values {\em and gradients} of model parameters. 
Compared to FL, D-SGD typically requires a tighter coordination among learning nodes and cannot handle 
their addition/removal, but the amount of data 
to transmit is lower; furthermore, unlike most FL variants, D-SGD does not require a learning server, 
and data can be exchanged directly between learning nodes, in a decentralized fashion. 
Due to its support for DNN splitting, D-SGD is a popular choice when the learning task involves nodes with limited capabilities, 
e.g., smart sensors~\cite{zhou2019edge}.  
D-SGD is said to implement {\em model parallelism}~\cite{krizhevsky2014one}, 
as different nodes run different parts of the same model.

A further, recently emerging alternative is represented by the {\em
split learning} (SL) paradigm~\cite{vepakomma2018split}. SL envisions
splitting the DNN layers in two parts: a {\em local} part, ran by each
node using its own local data, and a {\em global} part run at edge- or
cloud-based servers leveraging the intermediate outputs of all learning
nodes. It can be considered  
akin to D-SGD, in that the model is shared across multiple nodes, each
of which only implements a part of it. Compared to D-SGD, SL offers more
flexibility, e.g., by allowing to process data coming from different
sources through duplicate, parallel subsets of the DNN architecture.

In this work, we envision a new distributed supervised learning paradigm, 
called {\em flexible parallel learning (FPL)}, that allows for an increased flexibility in leveraging
 the resources available at Internet-of-Things (IoT) and mobile user devices. 
FPL achieves this goal by {\em combining  both data and model parallelism},
thus inherently bringing the benefits coming from both approaches.
Like D-SGD and SL, 
it enables splitting one model across multiple nodes, 
thereby associating different 
learning tasks to different nodes, 
and, like FL, it can leverage an arbitrary 
number of nodes performing the same learning task on their own local data:
knowledge of such a number is not needed at the time of designing the DNN architecture, so long as it remains constant for the duration of the training.
At the same time, FPL supports {\em partial replication}
of the model across multiple devices working on 
different data, thereby achieving indeed both model and data parallelism. 
Such higher flexibility is obtained by including an  additional layer   
in the DNN topology to accommodate  data 
coming from different sources. Importantly, such layer can also 
be effectively used to properly weigh input data depending on their quality.
To assess the impact of this additional layer, 
we  compare the FPL communication, computation, and energy costs  
against those  of FL and D-SGD, using state-of-the-art DNN architectures and large-scale, 
{\em de facto} standard datasets.

The remainder of the paper is organized as follows. 
\Sec{gdl} introduces the FPL paradigm, along with the main use cases it targets. 
Then \Sec{exp} and \Sec{results} describe the experiments we perform and the insights they provide. 
Finally, \Sec{conclusion} concludes the paper.

\begin{figure}
\centering
\includegraphics[width=.6\columnwidth]{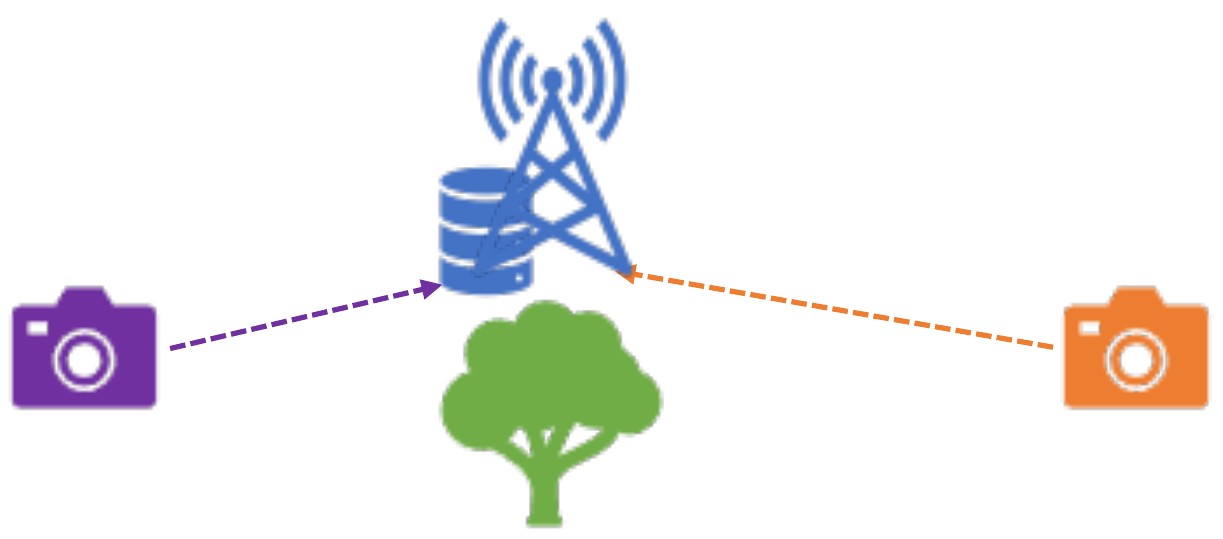}\\
\vspace{2mm}
\includegraphics[width=.6\columnwidth]{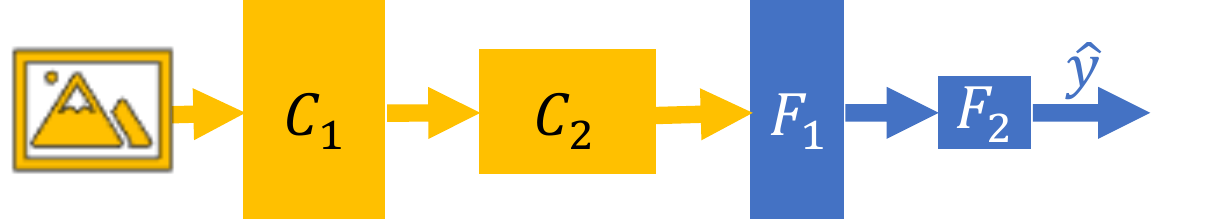}
\caption{
Top: example scenario, with two cameras observing the same object, capturing different views thereof. Bottom: example DNN architecture, taken from~\cite{caldas2018leaf}, for image classification.
\label{fig:scenario}
} 
\end{figure}

\section{The Flexible Parallel Learning Paradigm}
\label{sec:gdl}

{\bf Learning problem formulation.}
A distributed learning problem includes a set~$\mathcal{I}=\{1,\dots,I\}$ of {\em learning nodes}, all cooperating in order to optimize a model composed of {\em layers}~$\mathcal{L}=\{l_1,\dots,l_L\}$. Node~$i$ owns a local dataset~$\mathbf{X}_i$ and {\em local} parameters~$\mathbf{w}_i^m$, with~$m\in\mathcal{L}_i$ and $\mathcal{L}_i\subseteq\mathcal{L}$ denoting the layers running at node~$i$. The global model is given by the set of all parameters for all layers, with parameters of the same layer present at different nodes being averaged:
\begin{equation}
\label{eq:avg}
\mathbf{w}_l=\frac{1}{|\{i\in\mathcal{I}\colon l\in\mathcal{L}_i\}|}\sum_{i\in\mathcal{I}\colon l\in\mathcal{L}_i} \mathbf{w}_i^l.
\end{equation}
From the weights and the input data, it is possible to compute the output of the model~$\hat{\mathbf{y}}=g(\mathbf{X}_1,\dots,\mathbf{X}_I,\mathbf{w})$; as an example, in a classification problem, values in~$\hat{\mathbf{y}}$ represent the probabilities assigned to each class. The learning objective is to find the parameters~$\mathbf{w}^\star$ that minimize a loss function~$f(\hat{\mathbf{y}},\mathbf{y})$, where $\mathbf{y}$~are the ground-truth. 
For classification tasks, the most commonly-used loss function is the categorical cross-entropy
\begin{equation}
\label{eq:loss}
-\sum_{c=1}^C y_c\log\hat{y}_c,
\end{equation}
with~$c=1\dots C$ denoting the existing data classes.

The FPL paradigm targets IoT- and fog-based scenarios such as the one exemplified in \Fig{scenario}(top), 
where {\em multiple sources of data} and {\em multiple processing-capable devices} 
are available and can be leveraged to perform a common learning task, i.e., 
train a DNN such as the one exemplified in \Fig{scenario}(bottom). 
The high-level goal of FPL is to use all available data sources  (hence, achieving data parallelism), 
while distributing different parts of the DNN across the devices, without duplicating them unless 
needed (hence, achieving model parallelism). 
As an example, one may want to run the two convolutional layers, $C_1$ and~$C_2$, of \Fig{scenario}(bottom) 
separately at each of the two cameras, and then run {\em only one instance} of each of the fully-connected layers,
 $F_1$ and $F_2$, at an edge-based server.

\begin{figure}
\centering
\vspace{2mm} 
\includegraphics[width=1\columnwidth]{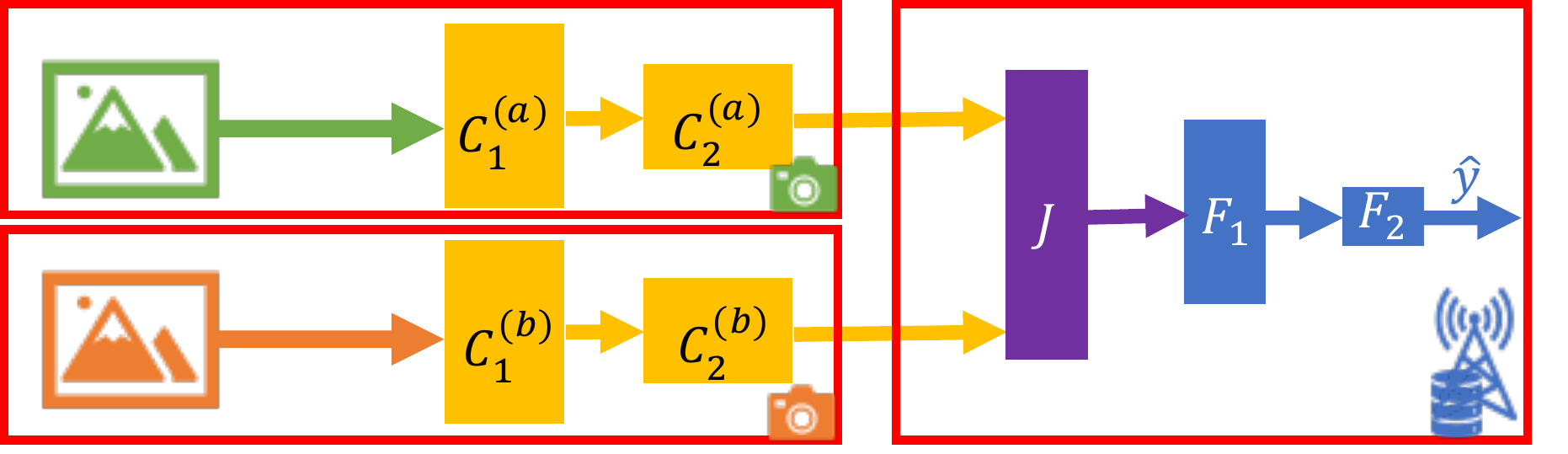}
\caption{
Example decisions made under the FPL paradigm. The two convolutional layers of the DNN in \Fig{scenario}(bottom) are duplicated, and an instance of each layer is ran at each of the two cameras in \Fig{scenario}(top). The outputs are then sent to the edge-based server, passed through the junction layer~$J$, and then fed to the rest of the original DNN.
\label{fig:solution}
} 
\end{figure}

More specifically, given a DNN architecture, the FPL paradigm operates as follows:
\begin{enumerate}
    \item it {\em identifies} the DNN layers that should be {\em replicated} at each learning node   
    to leverage multiple data sources;
    \item it {\em merges} the output of the above DNN layers through an appropriate junction layer; 
    \item it {\em distributes} the remaining DNN layers across the most suitable available nodes.
\end{enumerate}
With reference to the example  in \Fig{solution}, FPL creates two copies of $C_1$ and~$C_2$, 
i.e., one per data source. 
The outputs of the two $C_2$ copies, $C_2^{(a)}$ and~$C_2^{(b)}$, are then merged at the network edge 
through the junction layer $J$.
The input and output size of the junction layer match those of the preceding and succeeding layers; in the example of \Fig{solution}, $J$' input size will be the sum of $C_2^{(a)}$'s and~$C_2^{(b)}$'s output sizes, while $J$'s output size will be equal to the input size of~$F_1$.

We remark that a fundamental aspect of FPL is how it handles DNN merges.
One option would indeed be averaging the parameters, as in FL; however, such an approach is often suboptimal 
when the data sources are not equivalent, e.g., they observe different aspects of the 
phenomenon~\cite{noi-commag,jeong2018communication}.
D-SGD and SL opt instead for {\em statically} adapting the DNN architecture to the number of available data sources;
however, this (i) requires changing the size of~$F_1$ -- and potentially the whole DNN architecture --
when moving to a scenario with a different number of data sources,
and (ii) gives all data coming from all sources the same weight, 
regardless of their quality or importance. 
A better option is represented by the FPL paradigm, which combines the information coming from different 
data sources by including the junction layer 
in the DNN architecture, represented in purple in \Fig{solution}.  
The junction layer is fully-connected and  serves two purposes. 
The first is to adapt the number of parameters to the size of input and output layers; such a purpose is achieved by setting the input and output sizes of~$J$: with reference to \Fig{solution}, $J$'s input size is the sum of the output sizes of~$C_2^{(a)}$ and~$C_2^{(b)}$, while its output size is the same as the input size of~$F_1$. The second, and arguably most important, is to automatically learn {\em how to combine} the output of DNN layers running at different nodes. Indeed, the value of the parameters of layer~$J$ describe how information coming from different sources shall be processed, especially when they are of different quality.
Dataset quality itself is generally linked with whether a dataset over- or under-represents some data classes: datasets that do so are of poor quality, while i.i.d. datasets, adequately representing all types of data, are of high quality.

Dealing with low-quality, non-i.i.d. dataset is the focus of several FL studies, including~\cite{wang2019adaptive,infocom20-noniid,client-selection,li2018federated}, and the main strategy they use is assigning to learning nodes weights reflecting their data quality, thus, the contribution they can give to the learning process.
FPL achieves the same objective 
through the junction layer: the values of the parameters therein -- hence, the importance to assign to different data sources -- are found  
as a part of the DNN training process.
Indeed, the parameters of the junction layer are model parameters like all others, and are optimized through the same process -- forward- and back-passes, gradient optimization.

Replicating parts of a given DNN and merging them through a junction layer allows 
FPL to support both data and model parallelism, 
as well as different combinations thereof. 
If the scenario or the nature of data require so, FPL allows processing data coming from different 
sources in different ways, and distributing the necessary layers across various devices. At the same time, 
FPL supports scenarios where most of the processing is executed at the edge, and learning nodes only perform 
the operations necessary to reduce the quantity of data to transfer (e.g., running a convolutional layer).

{\bf Difference w.r.t. FL, D-SGD, and SL.}
FPL shares with the three distributed learning paradigms discussed earlier, namely, FL, D-SGD, and SL, the high-level goal of allowing a distributed set of nodes to cooperatively perform a learning task. There are, however, fundamental differences in the computations performed by nodes, and the data they exchange:
\begin{itemize}
    \item under FL, all nodes run the same model, and exchange, {\em after} one or more local epochs, the weights (parameters) of the whole model;
    \item under D-SGD and SL, each node runs a part, i.e., some layers, of the model, and node exchange gradient information {\em during} the forward- and backward-pass of {\em each} epoch.
\end{itemize}
Under FPL, nodes run a part of the model, similarly to D-SGD and SL, and
also communicate during each epoch. 
FPL is, however, superior to all alternatives due to the {\em
flexibility} afforded by the junction layer, during both the setup and
the training of the DNN. Thanks to the junction layer, the central
controller is able to adapt the DNN architecture to the number of data
sources at setup-time. Even more importantly, the junction layer
can learn, during the DNN training, how to best combine the information
coming from different data source: this includes how to deal with
different {\em types} of information, e.g., pictures and sensor
readings, but also giving a lower weight (or even a negative one) to
lower-quality information.

{\bf FPL scalability.}
The size of the junction layer added by FPL grows with the number of DNN branches to combine, which may pose a scalability problem. However, scalability is ensured by the fact that FPL allows flexibility as to {\em where} in the DNN the junction layer is placed. Indeed, the quantity of data exchanged between DNN layers decreases as we move closer to the solution, hence, moving the junction layer deeper into the DNN helps reducing the number of its parameters.

{\bf FPL training.}
A DNN built according to the FPL paradigm, i.e., including a junction layer, is nonetheless a DNN shared across multiple devices, as in the D-SGD or SL paradigms. It follows that such a DNN is trained with the same methodology and tools as in D-SGD and SL, i.e., with forward- and backward-passes within each epoch. Nodes exchange gradient information during each pass of each epoch, and parameters are optimized through distributed algorithms such as ADAM. Indeed, the FPL paradigm seeks to innovate how DNNs are {\em built and distributed} across devices, while leveraging existing training techniques.

{\bf Building DNN architectures with FPL.}
The flexibility of the FPL paradigm extends to its support for 
multiple decision-making strategies and algorithms: 
the paradigm itself does not mandate or require any specific strategy to choose 
which DNN layers to replicate and how to place them across the participating nodes.
Indeed, any of the existing approaches in the literature~\cite{jointframework,infocom20-fog} can be accommodated 
within the FPL paradigm.
However, the addition of a junction layer implies additional parameters to train  
and might lead to larger latency or processing power consumption. 
Assessing and quantifying such costs -- if any -- is indeed one of the main goal of our experiments. 

\begin{figure}
\centering
\includegraphics[width=.2\columnwidth]{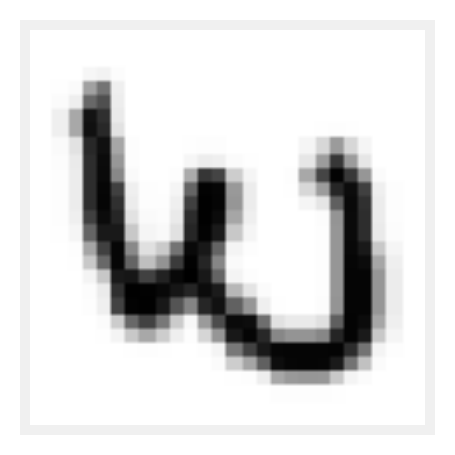}
\includegraphics[width=.2\columnwidth]{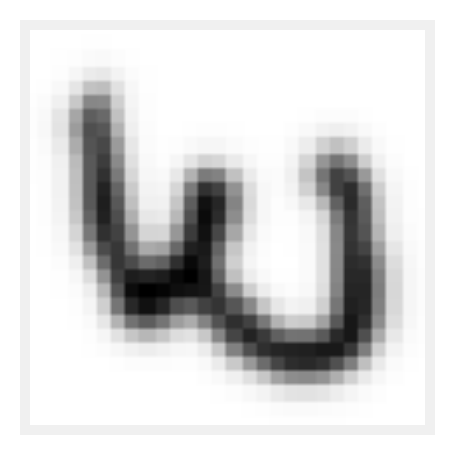}
\includegraphics[width=.2\columnwidth]{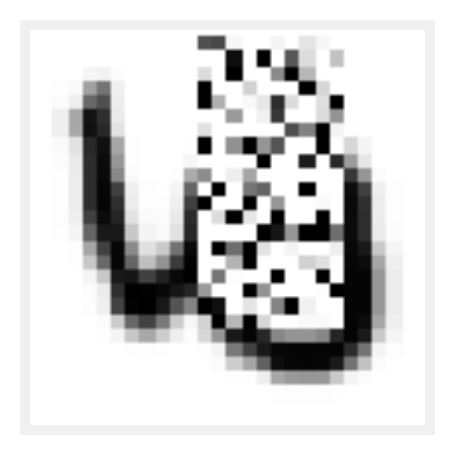}\\
\includegraphics[width=.2\columnwidth]{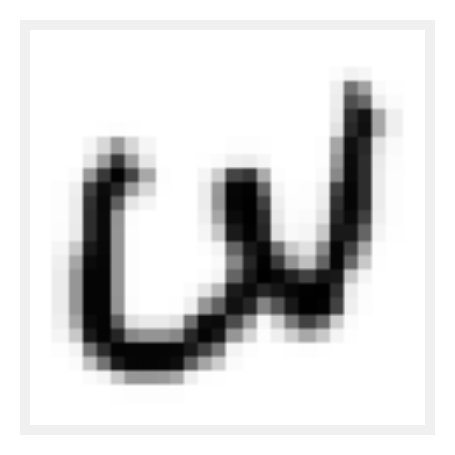}
\includegraphics[width=.2\columnwidth]{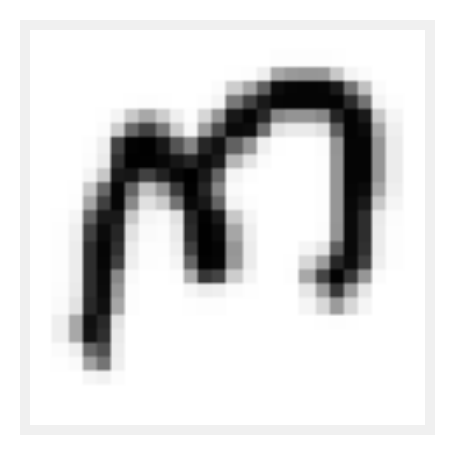}
\includegraphics[width=.2\columnwidth]{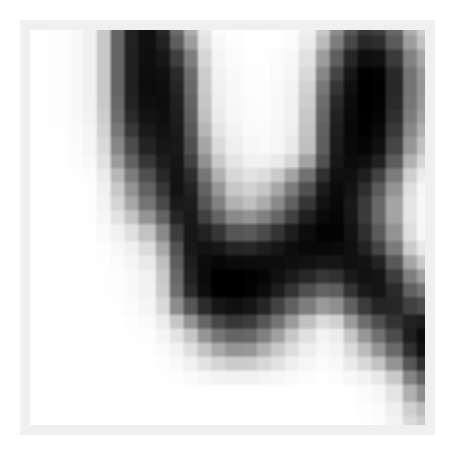}
\caption{
An example image from EMNIST: original (top left), blurred (top center), erased (top right), horizontally flipped (bottom left), vertically flipped (bottom center), cropped (bottom right).
\label{fig:emnist}
} 
\end{figure}

\section{Experiment design}
\label{sec:exp}

{\bf Channel and transmission model.}
Similar to~\cite{jointframework}, we assume that all nodes are equipped with cellular antennas and are covered by the same eNB, with the edge-based sever being co-located with an eNB itself. Nodes are randomly distributed in a 500~m-radius circular area around the eNB. The data rate achieved for transmissions from node~$i$ to node~$j$ is given by~\cite{jointframework}:
\begin{equation}
\label{eq:rate}
rB\log_2\mathds{E}_{h_i}\left(1+\frac{P_i h_i}{I+BN_0}\right),
\end{equation}
where~$r$ is the number of resource blocks (RBs) assigned to the communication, 
$B$~is the bandwidth of each RB, $P_i$ is the transmission power (30~dBm for the eNB, 10~dBm for the UEs), $N_0=-174$~dBm/Hz is the noise power spectral density. $h_i=o_i d_{ij}^{-2}$ is the channel gain, where $d_{ij}$~represents the distance between the terminals, and~$o_i$ the Rayleigh fading parameter. Consistently with LTE, we consider a 20~MHz bandwidth, divided into 100~RBs, assigned according to proportional-fair scheduling.

We compare the FPL paradigm performance and costs against state-of-the-art alternatives 
from the viewpoint of (a) learning accuracy, (b) model size, (c) learning time, (d) network overhead, 
and (e) energy consumption. 
We assess the performance of FPL and its alternatives over a classification task, 
based on the EMNIST dataset~\cite{cohen2017emnist}.

EMNIST has been introduced to provide a more challenging variant of the famous MNIST handwritten number dataset, 
and includes a total of 814,255~images belonging to 62~classes (26~uppercase and lowercase letters, plus ten digits). 
All images are in gray-scale and have a size of $28\times 28$~pixels, hence, they can be represented as~$28\times 28\times 1$ 
three-dimensional 
tensors.
The main reasons for using the EMNIST dataset are its relative simplicity and wide availability. 
Due to the former, we can compare the performance of FPL to its alternatives using a streamlined 
DNN architecture, as detailed next. Thanks to the latter, our results can be reproduced,  
generalized and compared against other existing works.

As shown in \Fig{scenario}(top), our goal is to assess the effect of FPL and 
its alternatives when dealing with different, 
partial views of the same phenomenon. 
To emulate this, we apply to the EMNIST 
one of the five transformation exemplified in \Fig{emnist}: Gaussian blur; random erasure; horizontal or vertical flip; random crop. 
Unless specified otherwise, to perform the classification task, we use the DNN architecture proposed in~\cite{caldas2018leaf} 
and represented in \Fig{scenario}(bottom), including two CNN layers (each followed by a max-pooling layer, 
not represented in the figure) and two fully-connected layers. 
The DNN, as well as the image transformations, is implemented in Python using the PyTorch framework. 
We selected PyTorch over the more popular TensorFlow framework due to the greater control the former affords 
over the manipulation of data, which simplifies implementing the newly proposed FPL paradigm.

We compare FPL against the following alternatives:

\noindent
{\em Split Learning}: we implement the ``vertical partitioned data'' variant of SL, introduced in~\cite[Sec.~2]{vepakomma2018split}: the convolutional layers are duplicated, with an instance thereof running at each data source. The input size of layer~$F_1$, which is the first non-duplicated layer, is adjusted so that the concatenated outputs of the duplicated layers can be fit therein.

\noindent
{\em Transfer images}: one single model is trained using 
all of the five image sources. 
    This entails transferring the images to the edge, thus incurring additional network overhead and delay; 
    
\noindent
    {\em Generalized FL} (gFL): models for 
each of the five image sources 
are separate, but some or all 
    their layers are 
    averaged at the end of each epoch. This reproduces and generalizes FL. 
Since the input datasets are not i.i.d., the FedProx strategy~\cite{li2018federated} is used {\em in lieu} of the more common FedAvg to combine the updates coming from different nodes. One averaging round is executed after each local computation one.

We run our experiments using a server with a $40$-core Intel Xeon E5-2690 v2 3.00\,GHz CPU, $64$~GB of memory, and equipped with a Tesla~K80 GPU.
We have opted not to use GPUs for training, in spite of the fact that PyTorch is able to exploit them, and 
doing so would have resulted in faster learning times. The main reason is that GPU usage is comparatively tricky to track, 
while -- via the \path{psutil} library -- it is possible to accurately measure the CPU consumption (hence, 
learning time and energy cost) 
associated with both FPL and the benchmark solutions.

\begin{figure}
\centering
\includegraphics[width=.29\textwidth]{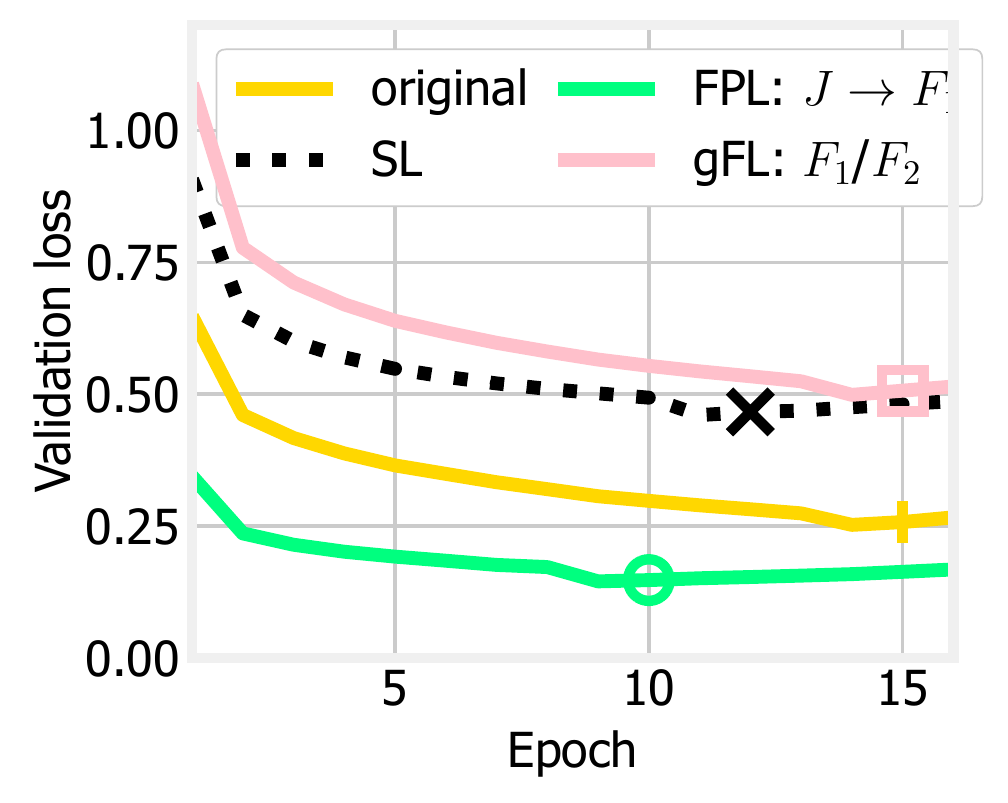}
\caption{
Validation loss of FPL and its alternatives. Markers indicate the iteration at which the loss reaches its minimum value, hence, training can be considered complete.
\label{fig:convergence}
} 
\end{figure}

\begin{figure*}
\vspace{1mm} 
\centering
\subfigure[\label{fig:accuracy}]{
    \includegraphics[width=.22\textwidth]{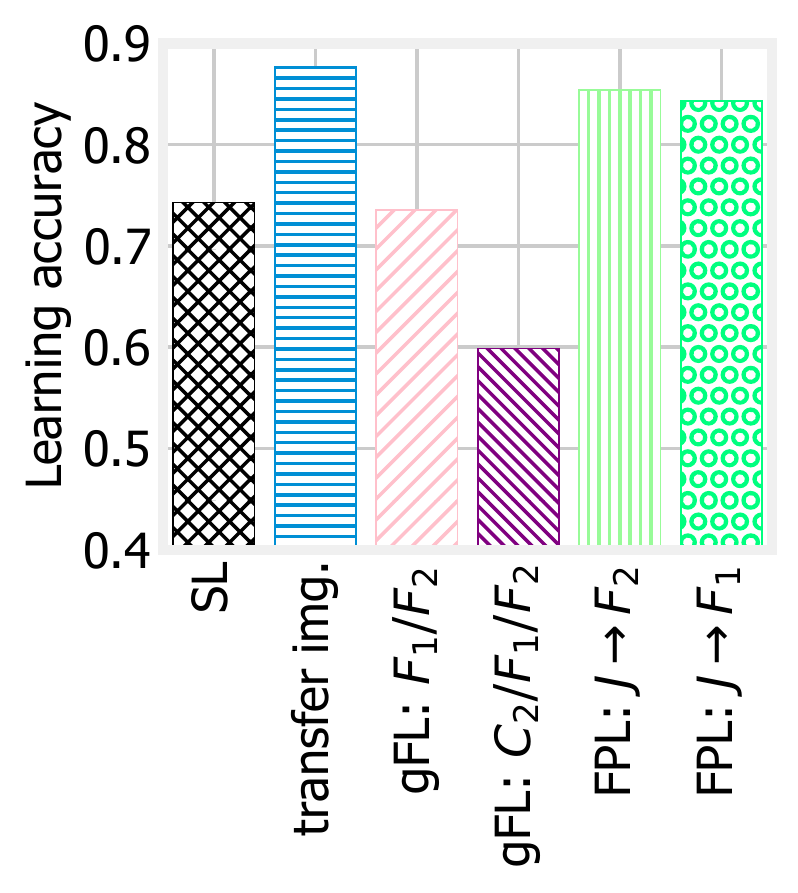}
} 
\hspace{-2mm}
\subfigure[\label{fig:size}]{
    \includegraphics[width=.22\textwidth]{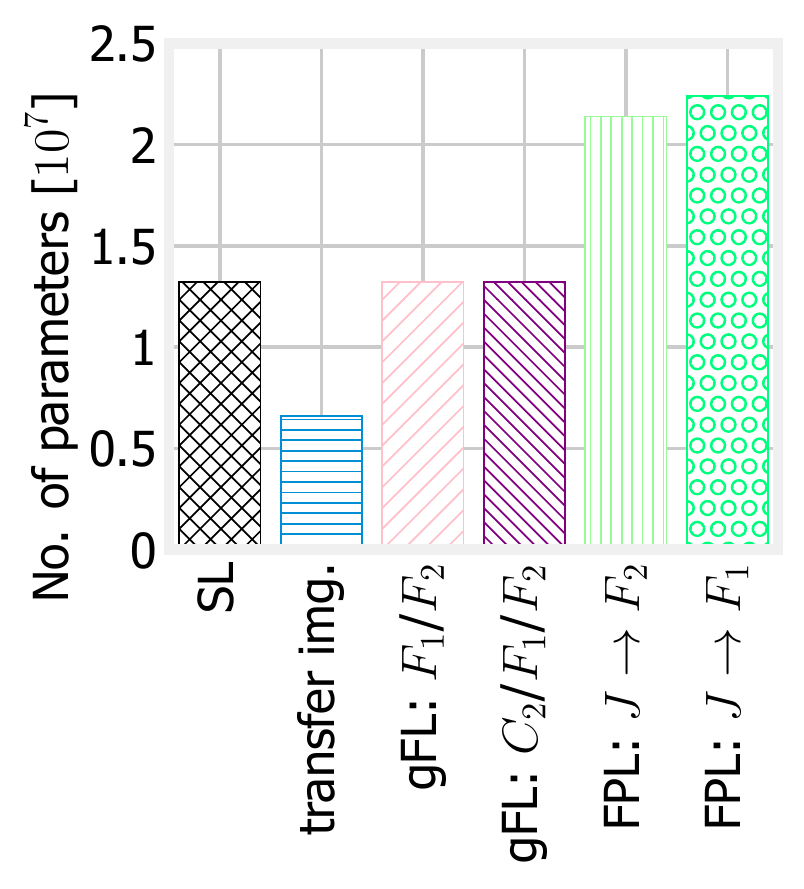}
} 
\hspace{-2mm}
\subfigure[\label{fig:time}]{
    \includegraphics[width=.22\textwidth]{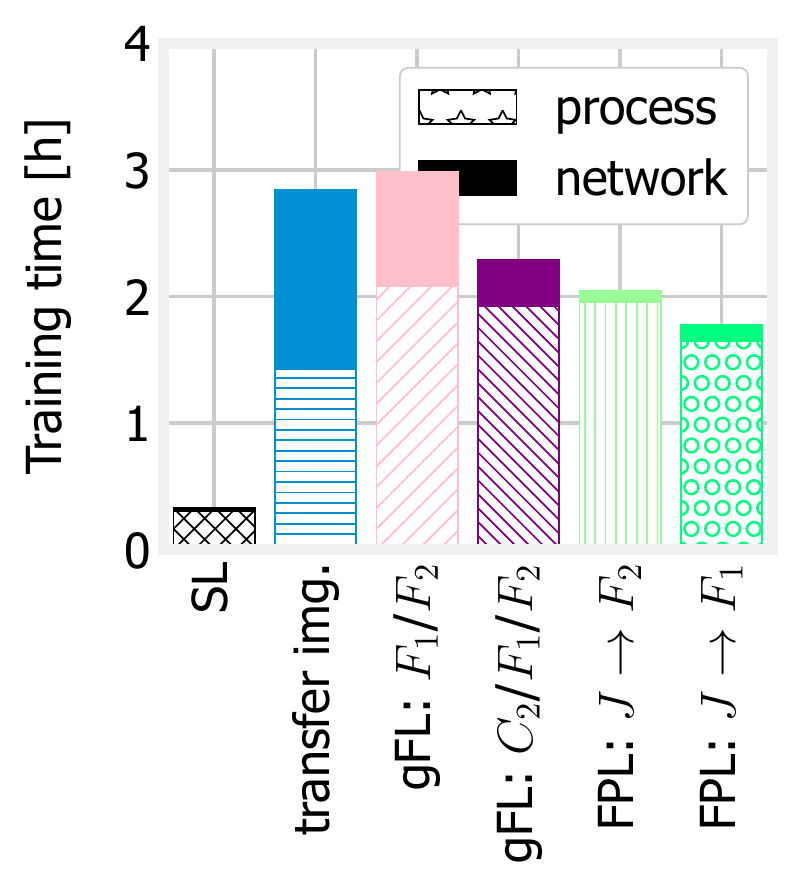}
} 
\hspace{-2mm}
\subfigure[\label{fig:network}]{
    \includegraphics[width=.22\textwidth]{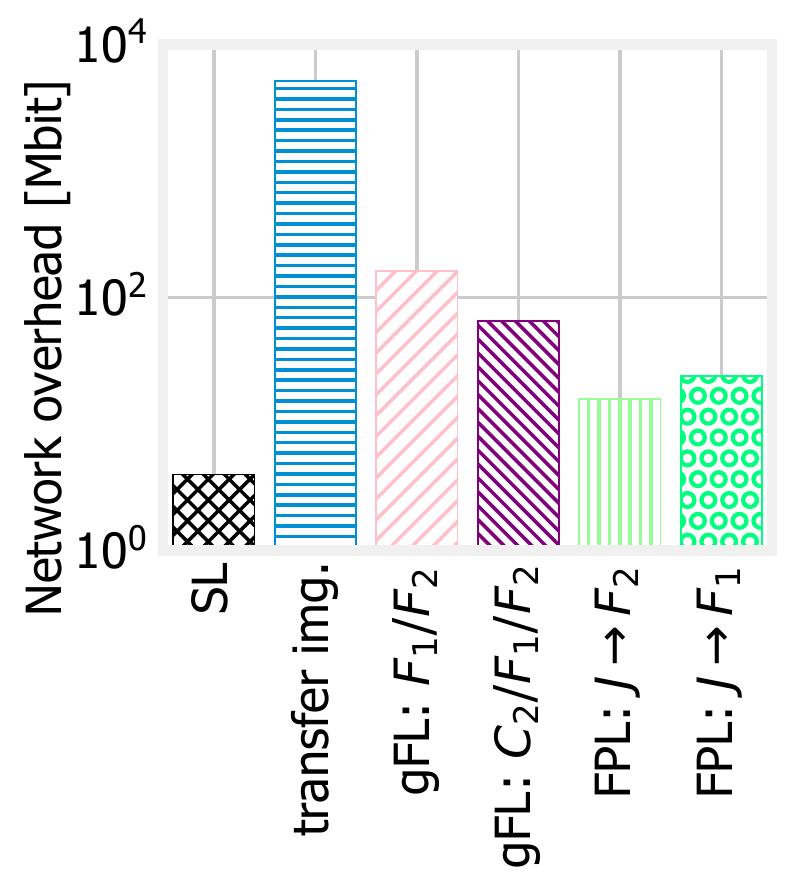}
} 
\caption{
Classification task over the EMNIST dataset: learning accuracy (a), number of model parameters (b), training time (c), and network overhead (d) under the FPL paradigm and its alternatives.
\label{fig:results}
} 
\end{figure*}

\section{Performance evaluation and discussion}
\label{sec:results}

The first aspect we are interested into is the convergence behavior of FPL and its alternatives, 
depicted in \Fig{convergence},
where $J\rightarrow F_1$ denotes that the junction 
layer in FPL is placed before layer~$F_1$ of the original DNN, i.e., 
between the second convolutional layer~$C_2$ and the first convolutional layer~$F_1$. 
For gFL, we list the layers that are averaged {\em \`{a} la} FL, e.g., $F_1/F_2$ 
indicates that the two fully-connected layers are averaged 
(while the convolutional ones are kept separate). 

The plot shows the value of the loss function \Eq{loss}, 
computed over the validation set at each iteration; convergence is achieved when the loss function 
starts increasing, which signals overfit. 
It is easy to observe how FPL has the fastest convergence, 
i.e., it takes fewer epochs to train than its alternatives. This translates into shorter training times, even if more operations (e.g., training the junction layer) must be performed at each epoch.

We next look at
the classification accuracy, summarized in \Fig{accuracy}. Similarly to \Fig{convergence}, for FPL we indicate the layer of the original DNN {\em before which} the junction layer is inserted, and for gFL we list the averaged layers.

As one might expect, transferring the images (bar with blue, horizontal line-pattern) 
results in the best performance; 
intuitively, this is because more data is used for training. The FPL paradigm (two rightmost bars with green patterns) 
yields the next-best performance; 
specifically, inserting the junction layer~$J$ just before the final fully-connected layer~$F_2$ is associated 
with the highest accuracy.
gFL (bars with pink/purple slanting-line patterns)
is associated with 
comparatively poor accuracy. Indeed, FL-like approaches are best suited when different 
learning nodes have access to similar images, e.g., the same view captured from different cameras.
If, however, the images are qualitatively different, e.g., flipped as in \Fig{emnist}, 
forcing the same model to process all images 
may be a suboptimal approach.
Using advanced strategies like FedProx~\cite{li2018federated} to combine the updates coming from different nodes does improve the performance compared to FedAvg, due to FedProx's ability to deal with non-i.i.d. data, but does not close the performance gap.
As for SL, it outperforms gFL, but yields a lower accuracy than FPL. 
Furthermore, it is important to stress that the configuration of the DNN in SL is tied to 
the number of data sources (five in our case). While FPL and gFL could accommodate 
any number of data sources, doing so in SL would require restructuring the whole DNN, 
which does not suit dynamic scenarios such as IoT- and fog-based ones.

\Fig{size} shows the size of models under the different paradigms, quantified through the number of their 
parameters. 
As one might expect, FPL is associated with a larger model size, due to the introduction of the junction 
layer~$J$. 
However, it is important to remark that
the increase in model size is moderate in both cases. 

It is more critical to assess whether the extra parameters introduced by FPL also result in longer learning times. 
As highlighted in \Fig{time}, FPL's learning times are comparable with those of its alternatives, 
actually shorter than those of gFL: 
indeed, the higher number of parameters to train is compensated by the fact that learning itself is 
more effective,
i.e., as per \Fig{convergence}, requires fewer epochs. From \Fig{time} it is also possible to observe how, unless we transfer all images to all learning nodes, processing represents the main contribution to the global training time.

\Tab{energy} summarizes the energy cost and carbon footprint of each approach, 
computed according to~\cite{lacoste2019quantifying} for
private servers running in northern Italy and supplied by the national energy provider Enel, which has a carbon efficiency of 0.243~kg/kWh according to \path{electricitymap.org}.
Consistently with \Fig{time}, it is clear that the FPL paradigm yields lower energy consumption and pollution.

Looking at \Fig{accuracy}--\Fig{time}, one may observe that transferring the images results in the highest 
accuracy, the smallest model size and a short training time, but \Fig{network}, depicting the network overhead, 
reminds why such a solution is not viable in virtually all scenarios. 
More importantly, the figure also highlights how FPL is associated with a 
very small network overhead -- lower than gFL approaches --, regardless of where the junction layer~$J$ 
is placed. 
This is due to the fact that, as highlighted in \Fig{approaches}, FL (unlike FPL) requires transmitting 
all model parameters. 
Also notice how the scale of \Fig{network} is logarithmic, and the advantage of FPL over FL-based approaches 
is almost one order of 
magnitude. 
Comparing \Fig{size} to \Fig{time} and \Fig{network}, it is also interesting to remark how 
the introduction of the junction layer~$J$ under then FPL paradigm mildly increases the model size, 
but does not result in longer training times or higher network overhead with respect to gFL.
Intuitively, 
there are more parameters but they are easier to train, and do not travel from a learning node to 
another.

\begin{table}
\caption{
Energy consumption and carbon footprint associated with FPL and its alternatives. Figures are computed 
according to~\cite{lacoste2019quantifying}
\label{tab:energy}
} 
\begin{tabularx}{1\columnwidth}{|X|r|r|}
\hline
Strategy & Energy [kWh] & Carbon [g CO$_2$ eq.] \\
\hline\hline
SL & 0.11 & 38.74 \\
\hline
Transfer images & 0.13 & 45.07 \\
\hline
gFL: $F_1$/$F_2$ & 0.23 & 77.97 \\
\hline
gFL: $C_2$/$F_1$/$F_2$ & 0.33 & 112.68 \\
\hline
FPL: $J\to F_2$ & 0.21 & 71.45 \\
\hline
FPL: $J\to F_1$  & 0.25 & 84.98 \\
\hline
\end{tabularx}
\end{table}

In summary, the FPL paradigm proves to be an attractive alternative to present-day approaches 
like FL or D-SGD. It provides better learning accuracy, allows for much higher flexibility and, unlike D-SGD, 
is a good match for dynamic scenarios -- at the price of a modest increase 
in model size and with comparable learning times and energy consumption.

\section{Conclusion}
\label{sec:conclusion}

To meet the needs of IoT- and fog-based environments, we have introduced a new distributed supervised learning paradigm 
called flexible parallel learning (FPL). FPL exhibits the advantages of both federated learning and distributed 
gradient descent, supports the dynamic addition and removal of learning nodes, 
and achieves both data and model parallelism. 
One of the most distinctive features of FPL is the introduction of a junction layer, through which 
data coming from different sources can be properly combined. 
We have evaluated the performance of FPL and its present-day alternatives over an image classification task, 
using the EMNIST dataset. 
The results highlight how FPL outperforms its alternatives in terms of learning accuracy,  
with comparable or lower network overhead and energy consumption.

Future work will focus on extending the performance evaluation of the FPL paradigm 
 considering the ability of the junction layer to differently weigh the input data, 
as well as additional datasets, e.g., CARS or ImageNet, and more complex DNN architectures, e.g., 
ResNet and VGG. Although the FPL paradigm itself works unmodified in these cases, 
the different sizes of the DNNs and the images they process may impact the relative performance 
of FPL and its alternative strategies.

\bibliographystyle{IEEEtran}
\bibliography{refs}%

\end{document}